\documentstyle[11pt,amssymb,epsfig,psfig]{article}
\def\beq{\begin{equation}}
\def\eeq{\end{equation}}

\def\text#1{\mbox{\scriptsize #1}}

\begin{document}

\title{Local Casimir energy for a wedge with a circular outer boundary}

\author{Amir H Rezaeian
 $^1$ \footnote{E-mail: Rezaeian@Theory.phy.umist.ac.uk}
 and Aram A Saharian  $^{2}$\footnote{E-mail: saharyan@www.physdep.r.am }  \\
 $^1$ Department of Physics, University of Manchester \\
 Institute of Science and Technology (UMIST), \\
 PO Box 88, Manchester, M60 1QD, UK \\
$^2$ Department of Physics, Yerevan State University \\
 1 Alex Manoogian St, 375049 Yerevan, Armenia }

\maketitle

\begin{abstract}
The local Casimir energy is investigated for a wedge with and
without a circular outer boundary due to the confinement of a
massless scalar field with general curvature coupling parameter
and satisfying the Dirichlet boundary conditions. Regularization
procedure is carried out making use of a variant of the
generalized Abel-Plana formula, previously established by one of
the authors. The surface divergences in the vacuum expectation
values of the energy density near the boundaries are considered.
The corresponding results can be applied to the cosmic strings.
\end{abstract}

\newpage

\section{Introduction}

There are two main mechanisms which contribute to the energy of
the vacuum: spontaneous symmetry breaking and the Casimir effect. The
Casimir effect is a direct consequence of the quantum field theory
and it is one of the most interesting macroscopic manifestations
of the nontrivial properties of the physical vacuum. Although the
Casimir effect has been extensively studied
\cite{plunien,cas1,milton,Bord01} there are still difficulties in
both its interpretation and renormalization \cite{milton,Seta00}.
Moreover, the absence of a complete renormalization procedure, in
practice, limits all calculations to the special case of
highly-symmetric boundary configurations (parallel plates, sphere,
cylinder) with a specific background metric. From this point of
view the wedge with a cylindrical outer boundary is an interesting
system, since the geometry is nontrivial and it includes two
dynamical parameters, radius and angle, for phenomenological
purposes. Due to the formal analogy that exists between a wedge
and a straight cosmic string, the corresponding results can be
applied to cosmic strings. Cosmic strings predicted in the
framework of various gauge theories with spontaneously broken
symmetries could have been created at cosmological phase
transitions in the early universe \cite{kibble,Vile94}. In the
case of a static and straightline cosmic string the outside
geometry is a locally flat conical spacetime with an angle
'deficit' $8\pi \mu $ and $\mu $ is the linear mass density of the
string. The superconducting strings form a particular subclass of
cosmic strings \cite{witten}. In the case of scalar field, it is
reasonable to impose a Dirichlet boundary condition at $r=a$
\cite{parker}. The investigation of the influence of the cosmic
strings on the behavior of quantized fields have been carried out
in \cite{cosmic}. In this paper we will study the Casimir energy
density for a massless scalar field with a general curvature
coupling parameter inside a wedge of opening angle $\phi _{0}$
with and without a cylindrical outer boundary assuming the
Dirichlet boundary conditions on the constraining surfaces. Some
most relevant investigations to the present paper are contained in
\cite{jphy,Deutsch,cas1,brevikI,brevikII} for a conformally
coupled scalar and electromagnetic fields in a four dimensional
spacetime. The total Casimir energy of a semi-circular infinite
cylindrical shell with perfectly conducting walls is considered in
\cite{nesterenko} by using the zeta function technique. Our method
here employs the mode summation and is based on a variant of the
generalized Abel-Plana formula \cite{Saha87} together with an
adequate cutoff function. This allows to extract from the vacuum
expectation value of the energy density the part due to a wedge
without outer cylindrical shell and to present the cylindrical
part in terms of the strongly convergent integrosum. We have
organized the paper as follows. The next section is devoted to the
consideration of the local Casimir energy for a massless scalar
field with a general curvature coupling inside a wedge. This
investigation generalizes the previously known result for a
conformally coupled scalar and is essential for the main purpose
of this paper in section \ref{sec:Wedgecyl}, where the vacuum
densities are considered for a wedge with the outer cylindrical
shell. A formula for the shell contribution to the vacuum energy
density is derived and the corresponding surface divergences are
investigated. Finally, the results are re-mentioned and discussed
in section \ref{sec:Conc}.

\section{Vacuum energy density inside a wedge for a scalar field with general
curvature coupling} \label{sec:Wedge}

In this section we will consider a scalar field $\varphi $, with
the curvature coupling parameter $\xi $, obeying Dirichlet
boundary condition on the boundary of the wedge-shaped region
formed by two plane boundaries intersecting at an arbitrary angle
$\phi _{0}$:
\begin{equation}
\phi (t,r,\phi =0,z_{1},\ldots ,z_{N})=\phi (t,r,\phi =\phi
_{0},z_{1},\ldots ,z_{N})=0,  \label{Dirbc}
\end{equation}
where we use cylindrical coordinates $(r,\phi ,z_{1},\ldots ,z_{N})$ in $%
D=N+2$ -- dimensional space. The corresponding field equation is in form
\begin{equation}
\left( \nabla ^{i}\nabla _{i}+\xi R\right) \varphi (x)=0,  \label{fieldeq}
\end{equation}
with $R$ being the scalar curvature for the background spacetime.
By using this equation, in the case of the flat background the
corresponding metric energy-momentum tensor may be presented in
the form
\begin{equation}
T_{ik}=\nabla _{i}\varphi \nabla _{k}\varphi +\left[ \left( \xi -\frac{1}{4}%
\right) g_{ik}\nabla ^{\mu }\nabla _{\mu }-\xi \nabla _{i}\nabla _{k}\right]
\varphi ^{2}.  \label{EMT}
\end{equation}
The vacuum expectation values for these quantities can be derived by
evaluating the mode sum
\begin{equation}
\langle 0|T_{ik}(x)|0\rangle =\sum_{{\mathbf{\alpha }}}T_{ik}\left\{ \varphi _{%
\mathbf{\alpha }}(x),\varphi _{{\mathbf{\alpha }}}^{\ast
}(x)\right\} , \label{vevEMT}
\end{equation}
where $\{\varphi _{{\mathbf{\alpha }}}(x)\}$ is a complete
orthonormal set of
positive frequency solutions to the field equation with quantum numbers $%
\alpha $, satisfying the corresponding boundary conditions. In the region
inside the wedge the corresponding eigenfunctions have the form
\begin{eqnarray}
\varphi _{\alpha }(x) &=&\frac{1}{(2\pi )^{N/2}}\left( \frac{\gamma }{\omega
\phi _{0}}\right) ^{1/2}J_{l}(\gamma r)\sin (l\phi )\exp \left( i{\mathbf{kr}}%
_{\parallel }-i\omega t\right) ,\quad \omega =\sqrt{\gamma ^{2}+k^{2}},\quad
l=\frac{n\pi }{\phi _{0}},  \label{eigfunc0} \\
\alpha  &=&(n,\gamma ,{\mathbf{k}}),\quad 0<\gamma <\infty ,\quad -\infty <%
k_j<\infty ,\quad n=1,2,\cdots ,  \nonumber\end{eqnarray} where
${\mathbf{r}}_{\parallel }=(z_{1},\ldots ,z_{N})$, and $J_{l}(z)$
is the Bessel function. Substituting the eigenfunctions into
(\ref{vevEMT}) for the corresponding vacuum energy density one
finds
\begin{equation}
\langle 0_{w}|T_{00}|0_{w}\rangle =\frac{1}{2(2\pi )^{N}\phi _{0}}%
\sum_{n=1}^{\infty }\int_{0}^{\infty }d\gamma \int d^{N}{\mathbf{k}}\frac{%
\gamma }{\omega }\left[ A_{l}^{(-)}\left( \gamma
,{\mathbf{k}},r\right) -A_{l}^{(+)}\left( \gamma
,{\mathbf{k}},r\right) \cos (2l\phi )\right] \label{T00}
\end{equation}
with $|0_{w}\rangle $ being the amplitude for the corresponding
vacuum inside a wedge, and we use the notations
\begin{equation}
A_{l}^{(\pm )}\left( \gamma ,{\mathbf{k}},r\right) =\omega
^{2}J_{l}^{2}(\gamma r)+\left( 2\xi -\frac{1}{2}\right) \gamma
^{2}\left[ \left( 1\pm \frac{l^{2}}{\gamma ^{2}r^{2}}\right)
J_{l}^{2}(\gamma r)-J_{l}^{^{\prime }2}(\gamma r)\right] .
\label{Anpm}
\end{equation}
To evaluate expression (\ref{T00}) firstly we integrate over
${\mathbf{k}}$ by using the formula
\begin{equation}
\int \frac{k^{s}d^{N}{\mathbf{k}}}{\sqrt{k^{2}+\gamma ^{2}}}=\gamma ^{N+s-1}%
\frac{\pi ^{(N-1)/2}}{\Gamma (N/2)}\Gamma \left( -\frac{N+s-1}{2}\right)
\Gamma \left( \frac{N+s}{2}\right) ,  \label{intk1}
\end{equation}
and next we integrate over $\gamma $ on the base of the formula
\begin{equation}
\int_{0}^{\infty }d\gamma \,\gamma ^{\alpha -1}J_{l}^{2}(\gamma r)=\frac{%
r^{-\alpha }}{2\sqrt{\pi }}\frac{\Gamma \left( \frac{1-\alpha }{2}\right)
\Gamma \left( l+\alpha /2\right) }{\Gamma \left( 1-\alpha /2\right) \Gamma
\left( l+1-\alpha /2\right) }.  \label{intgam1}
\end{equation}
The integral involving the derivative of the Bessel function can be
evaluated by using the relation
\begin{equation}
\left( 1-\frac{l^{2}}{\gamma ^{2}r^{2}}\right) J_{l}^{2}(\gamma
r)-J_{l}^{^{\prime }2}(\gamma r)=-\frac{1}{2\gamma ^{2}}\left( \frac{d^{2}}{%
dr^{2}}+\frac{1}{r}\frac{d}{dr}\right) J_{l}^{2}(\gamma r).  \label{Besrel1}
\end{equation}
As a result for the separate terms in (\ref{T00}) one finds
\begin{eqnarray}
\int_{0}^{\infty }d\gamma \int d^{N}{\mathbf{k}}\frac{\gamma }{\omega }%
A_{l}^{(-)}\left( \gamma ,{\mathbf{k}},r\right)  &=&\frac{\pi ^{N/2-1}}{%
4r^{N+3}}\Gamma \left( -N/2-1\right) \frac{\Gamma \left( l+\frac{N+1}{2}%
\right) }{\Gamma \left( l-\frac{N-1}{2}\right) }  \label{sum1An1} \\
&&\times \left[ (N+1)^{2}(N+2)(\xi -\xi _{c})-l^{2}\right]
\nonumber\end{eqnarray}
\begin{eqnarray}
\int_{0}^{\infty }d\gamma \int d^{N}{\mathbf{k}}\frac{\gamma }{\omega }%
A_{l}^{(+)}\left( \gamma ,{\mathbf{k}},r\right) \cos (2l\phi )
&=&\frac{\pi ^{N/2-1}}{r^{N+3}}(N+2)(\xi -\xi _{c})\Gamma \left(
-N/2-1\right)
\label{sum2An} \\
&&\times \left[ \frac{(N+1)^{2}}{4}-l^{2}\right] \frac{\Gamma \left( l+\frac{%
N+1}{2}\right) }{\Gamma \left( l-\frac{N-1}{2}\right) }\cos
(2l\phi ), \nonumber
\end{eqnarray}
where $\xi _{c}=(D-1)/(4D)$ is the curvature coupling parameter
for the conformal case. To obtain the regularized value of the
energy density we have to subtract from (\ref{T00}) the
corresponding quantity for the Minkowski vacuum without
boundaries. The latter can be found in a similar way. The
corresponding eigenfunctions are in form
\begin{equation}
\varphi _{\alpha }^{(M)}(x)=\frac{1}{(2\pi )^{(N+1)/2}}\left( \frac{\gamma }{%
2\omega }\right) ^{1/2}J_{n}(\gamma r)\exp \left( in\phi +i{\mathbf{kr}}%
_{\parallel }-i\omega t\right) ,\quad n=\cdots -2,-1,0,1,2\ldots .
\label{Minkmodes}
\end{equation}
Now on the base of formula (\ref{vevEMT}) one finds
\begin{equation}
\langle 0_{M}|T_{00}(x)|0_{M}\rangle =\frac{1}{(2\pi )^{N+1}}%
\sum_{n=0}^{\infty }{}^{\prime }\int_{0}^{\infty }d\gamma \int
d^{N} {\mathbf{k}}\frac{\gamma }{\omega }A_{n}^{(-)}\left( \gamma
,{\mathbf{k}},r\right) , \label{T00Mink}
\end{equation}
where $|0_{M}\rangle $ denotes the Minkowski vacuum, and the prime
means that in the sum over $n$ the term with $n=0$ is taken with
weight $1/2$. Taking into account that $\sum_{n=0}^{\infty
}{}^{\prime }J_{n}^{2}(\gamma r)=1/2$ it can be easily seen that
expression (\ref{T00Mink}) is presented in the standard form
\begin{equation}
\langle 0_{M}|T_{00}(x)|0_{M}\rangle =\int \frac{d^{D}{\mathbf{k}}}{(2\pi )^{D}%
}\frac{\omega }{2}.  \label{T00Minkst}
\end{equation}
After integrating over ${\mathbf{k}}$ and $\gamma $ in
(\ref{T00Mink}) we receive
\begin{equation}
\langle 0_{M}|T_{00}(x)|0_{M}\rangle =\frac{\Gamma \left( -N/2-1\right) }{%
2^{N+3}\pi ^{N/2+2}r^{N+3}}\sum_{n=0}^{\infty }{}^{\prime }\frac{\Gamma
\left( n+\frac{N+1}{2}\right) }{\Gamma \left( n-\frac{N-1}{2}\right) }\left[
(N+1)^{2}(N+2)(\xi -\xi _{c})-n^{2}\right] .  \label{T00Mink2}
\end{equation}
Let us consider the case $N=1$. The sums over $n$ are evaluated by
using the Riemann zeta function $\zeta (s)=\sum_{n=1}^{\infty
}n^{-s}$, $\zeta (-1)=-1/12$, $\zeta (-3)=1/120$, and the formula
\begin{equation}
\sum_{n=1}^{\infty }n^{p}\cos (\beta n)=\frac{1}{2}\sin \frac{\pi p}{2}\frac{%
d^{p}}{d\beta ^{p}}\cot (\beta /2),  \label{geo}
\end{equation}
valid for an odd $p$. For the regularized vacuum energy density this leads
\begin{eqnarray}
\langle T_{00}\rangle _{\mathrm{reg}} &=&\langle 0_{w}|T_{00}|0_{w}\rangle
-\langle 0_{M}|T_{00}|0_{M}\rangle =-\frac{(\pi /\phi _{0})^{4}-1}{1440\pi
^{2}r^{4}}-(\xi -\xi _{c})\frac{(\pi /\phi _{0})^{2}-1}{12\pi ^{2}r^{4}}+
\nonumber \\
&&+\frac{\xi -\xi _{c}}{4\phi _{0}^{2}r^{4}}\csc ^{2}(\pi \phi /\phi _{0})%
\left[ \frac{\pi ^{2}}{2\phi _{0}^{2}}\left( 3\csc ^{2}(\pi \phi
/\phi _{0})-2\right) +1\right] ,  \label{18}
\end{eqnarray}
where $\xi _{c}=1/6$ is the curvature coupling parameter for the
conformal case in $D=3$. In formula (\ref{18}) the first two terms
on the right come from the summand with $A_{l}^{(-)}\left( \gamma
,{\mathbf{k}},r\right) $ in
(\ref{T00}), and the last term comes from the summand with $%
A_{l}^{(+)}\left( \gamma ,{\mathbf{k}},r\right) $. For $\phi
_{0}=\pi $ it can be easily seen that expression (\ref{18})
coincides with the well-known result for the Casimir energy
density in the case of a single plate geometry. For the
conformally coupled case from (\ref{18}) we obtain the standard
result previously derived in Refs. \cite{Deutsch},\cite{kennedy}.
It is of interest to note that in this case the renormalized
energy density is angle independent and is finite on the sides of
the wedge. It diverges on the edge as $r^{-4}$ , when
$r\rightarrow 0$ (as $r^{-(D+1)}$ for a $D$ -- dimensional space).
For a non--conformally coupled scalar the energy density is angle
dependent and contains surface divergences on the wedge sides.
Note that for a minimally coupled scalar ($\xi =0$) the
regularized energy density (\ref{18}) is negative everywhere,
$\langle T_{00}\rangle _{\mathrm{reg}}|_{\xi =0}<0$.

\section{Vacuum energy density inside a wedge with circular boundary}
\label{sec:Wedgecyl}

Having investigated the vacuum energy density inside a wedge we now turn to
the case with additional circular boundary with radius $a$. The boundary
conditions are Dirichlet ones:
\begin{eqnarray}
&&\varphi (t,r,\phi =0,z_{1},\ldots ,z_{N})=\varphi (t,r,\phi =\phi
_{0},z_{1},\ldots ,z_{N})=0,  \label{bcond2} \\
&&\varphi (t,r=a,\phi ,z_{1},\ldots ,z_{N})=0.
\nonumber\end{eqnarray} The eigenfunctions satisfying these
boundary conditions are of the form
\begin{equation}
\varphi _{\alpha }(x)=\beta _{\alpha }J_{l}(\gamma r)\sin (l\phi
)\exp \left( i{\mathbf{kr}}_{\parallel }-i\omega t\right) ,
\label{eigfunccirc}
\end{equation}
where the notations are the same as in (\ref{eigfunc0}). The
normalization coefficient $\beta _{\alpha }$ is determined from
the standard Klein-Gordon scalar product and is equal to
\begin{equation}
\beta _{\alpha }^{2}=\frac{2}{(2\pi )^{N}\omega \phi _{0}a^{2}J_{l}^{\prime
2}(\gamma a)}.  \label{betalf}
\end{equation}
The eigenvalues for the quantum number $\gamma $ are quantized by the
boundary condition (\ref{bcond2}) on the cylinder surface $r=a$. From this
condition it follows that the possible values of $\gamma $ are equal to
\begin{equation}
\gamma =\lambda _{n,j}/a,\quad j=1,2,\cdots ,  \label{ganval}
\end{equation}
where $\lambda _{n,j}$ are positive zeros of the Bessel function, $%
J_{l}(\lambda _{n,j})=0$, arranged in ascending order, $\lambda
_{n,j}<\lambda _{n,j+1}$. Substituting the eigenfunctions (\ref{eigfunccirc}%
) into mode sum formula (\ref{vevEMT}) for the corresponding energy density
one finds
\begin{equation}
\langle 0|T_{00}|0\rangle =\frac{1}{2}\sum_{n=1}^{\infty
}\sum_{j=1}^{\infty }\int d^{N} {\mathbf{k}}\beta _{\alpha
}^{2}\left[ A_{l}^{(-)}\left( \lambda _{n,j}/a,
{\mathbf{k}},r\right) -A_{l}^{(+)}\left( \lambda
_{n,j}/a,{\mathbf{k}},r\right) \cos (2l\phi )\right] ,
\label{T00circ}
\end{equation}
with a set of quantum numbers $\alpha =(n,j,{\mathbf{k}})$. The
vacuum expectation value (\ref{T00circ}) is divergent. To make it
finite we introduce the cutoff function $\psi _{\mu }(\gamma )$
satisfying the condition $\psi _{\mu }\rightarrow 1$, $\mu
\rightarrow 0$. To extract the divergent part we will apply to the
sum over $j$ the summation formula \cite{Saha87}
\begin{eqnarray}
\sum_{j=1}^{\infty }\frac{f(\lambda _{n,j})}{\lambda _{n,j}J_{l}^{\prime
2}(\lambda _{n,j})\sqrt{\lambda _{n,j}^{2}+c^{2}}} &=&\frac{1}{2}%
\int_{0}^{\infty }\frac{f(z)}{\sqrt{z^{2}+c^{2}}}dz+\frac{\pi
}{4}{\mathrm{Res}}_{z=0}\left[
\frac{f(z)Y_{l}(z)}{\sqrt{z^{2}+c^{2}}J_{l}(z)}\right] -
\nonumber\\
&&-\frac{1}{2\pi }\int_{0}^{c}dz\,\frac{K_{l}(z)}{I_{l}(z)}\frac{e^{-l\pi
i}f(ze^{\pi i/2})+e^{l\pi i}f(ze^{-\pi i/2})}{\sqrt{c^{2}-z^{2}}}  \nonumber
\label{sumform1AP} \\
&&+\frac{i}{2\pi }\int_{c}^{\infty }dz\,\frac{K_{l}(z)}{I_{l}(z)}\frac{%
e^{-l\pi i}f(ze^{\pi i/2})-e^{l\pi i}f(ze^{-\pi i/2})}{\sqrt{z^{2}-c^{2}}},
\end{eqnarray}
where $Y_{l}(z)$ is the Neumann function, and $I_{l}(z)$, $K_{l}(z)$ are
Bessel modified functions. This formula is valid for functions $f(z)$
satisfying the conditions
\begin{equation}
|f(z)|<\epsilon (x)e^{c_{1}|y|},\quad z=x+iy,\quad c_{1}<2,
\label{sumformcond1}
\end{equation}
\begin{equation}
f(z)=o(z^{2|l|-1}),\quad z\rightarrow 0,  \label{sumformcond2}
\end{equation}
where $\epsilon (x)\rightarrow 0$ for $x\rightarrow \infty $.

To evaluate the sum over $j$ in (\ref{T00circ}) now as a function
$f(z)$ we choose
\begin{equation}
f(z)=z\left[ A_{l}^{(-)}\left( z/a,{\mathbf{k}},r\right)
-A_{l}^{(+)}\left( z/a,{\mathbf{k}},r\right) \cos (2l\phi )\right]
\psi _{\mu }(z/a), \label{ftosum}
\end{equation}
assuming a class of cutoff functions for which $f(z)$ satisfies
conditions (\ref{sumformcond1}) and (\ref{sumformcond2}) uniformly
with respect to the cutoff parameter $\mu $. It can be seen that
the contribution of the first integral on the right of \ formula
(\ref{sumform1AP}) to the energy density (\ref{T00circ})
corresponds to the energy density for the case of a wedge without
outer cylindrical boundary. This quantity was investigated in the
previous section. The second and third integrals on the right-hand
side of (\ref{sumform1AP}) are finite in the limit $\mu
\rightarrow 0$. Using the standard formulae for the Bessel
function we see that the subintegrand of the second integral is
proportional to $\psi _{\mu }(iz/a)-\psi _{\mu }(-iz/a)$.
Consequently, after removing the cutoff the contribution of this
integral will be zero. Hence, omitting this integral and removing
the cutoff for the vacuum energy density we obtain
\begin{eqnarray}
\langle 0|T_{00}|0\rangle &=&\langle 0_{w}|T_{00}|0_{w}\rangle -\frac{1}{%
2^{N-1}\pi ^{N+1}\phi _{0}a}\sum_{n=1}^{\infty }\int d^{N}{\mathbf{k}}%
\int_{ka}^{\infty }\frac{zdz}{\sqrt{z^{2}-k^{2}a^{2}}}\frac{K_{l}(z)}{%
I_{l}(z)}  \nonumber \label{T00circ10} \\
&&\times \left\{ \left[ k^{2}I_{l}^{2}(zx)-\left( 2\xi -\frac{1}{2}\right)
\frac{z^{2}}{a^{2}}\left( I_{l}^{\prime 2}(zx)+I_{l}^{2}(zx)\frac{2\xi +1/2}{%
2\xi -1/2}\right) \right] \sin ^{2}(l\phi )-\right.  \nonumber\\
&&-\left. \left( 2\xi -\frac{1}{2}\right) \frac{l^{2}I_{l}^{2}(zx)}{%
a^{2}x^{2}}\cos ^{2}(l\phi )\right\} ,\quad x=\frac{r}{a}.
\end{eqnarray}
The integration over ${\mathbf{k}}$ can be done using the formula
\cite{saharianI}
\begin{equation}\label{intk}
  \int d^{N}{\mathbf{k}}\int _{k}^{\infty
  }\frac{k^sg(z)dz}{\sqrt{z^2-k^2}}=\frac{\pi ^{N/2}}{\Gamma
  (N/2)}B\left( \frac{N+s}{2},\frac{1}{2}\right)\int _{0}^{\infty
  }dz z^{N+s-1}g(z),
\end{equation}
where $B(x,y)$ is the Euler beta function. As a result for the
vacuum energy density (\ref{T00circ10}) one obtains
\begin{equation}
\langle 0|T_{00}|0\rangle =\langle 0_{w}|T_{00}|0_{w}\rangle +\langle
T_{00}\rangle _{c},  \label{T00circ2}
\end{equation}
where $\langle 0_{w}|T_{00}|0_{w}\rangle $ is the vacuum energy
density for a wedge without outer cylindrical boundary, and the
term
\begin{equation}
\langle T_{00}\rangle _{c}=\frac{4\xi -1}{2^{D-1}\pi ^{(D-1)/2}\phi
_{0}a^{D+1}\Gamma ((D-1)/2)}\sum_{n=1}^{\infty }\int_{0}^{\infty }dz\,z^{D}%
\frac{K_{l}(z)}{I_{l}(z)}\left[ B_{l}^{(-)}\left( zx\right)
-B_{l}^{(+)}\left( zx\right) \cos (2l\phi )\right]  \label{T00boundc}
\end{equation}
is the contribution due to the presence of the cylindrical shell
at $r=a$. Here $l$ is defined in accordance with (\ref{eigfunc0})
and we have introduced the notations
\begin{equation}
B_{l}^{(\pm )}\left( y\right) =I_{l}^{\prime 2}(y)+I_{l}^{2}(y)\left[ 1+%
\frac{2}{(D-1)(4\xi -1)}\mp \frac{l^{2}}{y^{2}}\right] .  \label{Blpm}
\end{equation}
In accordance with the problem symmetry,  expression
(\ref{T00boundc}) is invariant under the replacement $\phi \to
\phi _0 -\phi $. Aiming to compare with the result for the energy
density of a cylindrical shell with the radius $a$ let us write
down the corresponding formula, which is obtained from the general
result of \cite{saharianI} and has the form
\begin{equation}\label{cyldens}
 \langle T_{00}\rangle _{{\mathrm{reg}}}^{{\mathrm{cyl}}}=
\frac{4\xi -1}{2^{D}\pi ^{(D+1)/2}a^{D+1}\Gamma ((D-1)/2)}
\sum_{n=-\infty }^{+\infty }\int_{0}^{\infty }dz\,z^{D}%
\frac{K_{n}(z)}{I_{n}(z)}B_{n}^{(-)}\left( zx\right) ,
\label{purecyl}
\end{equation}
with the same notations as in (\ref{T00boundc}).

For $0<r<a$ the cylindrical part (\ref{T00boundc}) is finite for
all values $0\leq \phi \leq \phi _{0}$, including the wedge sides.
The divergences on these sides are included in the first term on
the right-hand side of (\ref{T00circ2}), corresponding to the case
without circular boundary. At the edge $r=0$ the boundary part
(\ref{T00boundc}) vanishes for $0<\phi _0<\pi $, is equal to
\begin{equation}\label{boundr0}
  \langle T_{00}\rangle _{0}|_{r=0}=\frac{4\xi -1}{2^{D}\pi ^{(D+1)/2}
  \Gamma ((D-1)/2)a^{D+1}}\int _{0}^{\infty }dz
  z^{D}\frac{K_1(z)}{I_1(z)}, \quad \phi _0=\pi
\end{equation}
for $\phi _0=\pi $, and diverges as $r^{-2(1-\pi /\phi _0)}$ for
$\phi _0>\pi $.

The boundary part $\left\langle T_{00}\right\rangle _{c}$ diverges
on the cylindrical surface $r=a$ as well. To investigate the
corresponding behavior near this surface let us consider the
integrosum in (\ref{T00boundc}), introducing a new integration
variable $z\rightarrow lz$:
\begin{equation}
{\mathcal{I}}=\sum_{n=1}^{\infty }l^{D+1}\int_{0}^{\infty }dz\,z^{D}\frac{%
K_{l}(lz)}{I_{l}(lz)}\left[ B_{l}^{(-)}\left( lzx\right) -B_{l}^{(+)}\left(
lzx\right) \cos (2l\phi )\right] .  \label{integrosum}
\end{equation}
By taking into account that near the surface $r=a$ main
contribution comes from the large values of $l$ we can replace the
Bessel modified functions by their uniform asymptotic expansions
for large values of the order (see, for instance \cite{hand}). In
the leading order over $1/l$ one finds
\begin{equation}
{\mathcal{I}}=\frac{1}{2x^{2}}\sum_{n=1}^{\infty
}l^{D}\int_{0}^{\infty }dz\,z^{D-2}(1+z^{2}x^{2})^{1/2}e^{-2l[\eta
(z)-\eta (zx)]}\left[ b^{(-)}(zx)-b^{(+)}(zx)\cos (2l\phi )\right]
,  \label{intsum1}
\end{equation}
where
\begin{eqnarray}
b^{(\pm )}(y) &=&1+\frac{1}{1+y^{2}}\left[ y^{2}\mp 1+\frac{2y^{2}}{%
(D-1)(4\zeta -1)}\right] ,  \label{bepm} \\
\eta (z) &=&\sqrt{1+z^{2}}+\ln \frac{z}{1+\sqrt{1+z^{2}}}.
\end{eqnarray}
Near the cylindrical boundary in the leading order over $1-x$ from
(\ref{intsum1}) we obtain
\begin{equation}
{\mathcal{I}}=\frac{1}{2}\left( \frac{\pi }{\phi _{0}}\right)
^{D}\sum_{n=1}^{\infty }n^{D}\int_{0}^{\infty
}dz\,z^{D-2}(1+z^{2})^{1/2}e^{-2n\pi (1-x)\sqrt{1+z^{2}}/\phi
_{0}}\left[ b^{(-)}(z)-b^{(+)}(z)\cos (2l\phi )\right] .
\label{intsum3}
\end{equation}
On the wedge sides $\cos (2l\phi )=1$ and this yields
\begin{equation}
{\mathcal{I}}=\left( \frac{\pi }{\phi _{0}}\right)
^{D}\sum_{n=1}^{\infty
}n^{D}\int_{0}^{\infty }\frac{dz\,z^{D-2}}{(1+z^{2})^{1/2}}e^{-2n\pi (1-x)%
\sqrt{1+z^{2}}/\phi _{0}},\quad \phi =0,\quad \phi =\phi _{0}.
\label{intsum4}
\end{equation}
Summing over $n$ by using
\begin{equation}
\sum_{n=1}^{\infty }n^{D}e^{-\alpha n}=(-1)^{D}\frac{d^{D}}{d\alpha ^{D}}%
\frac{1}{e^{\alpha }-1}\approx \frac{\Gamma (D+1)}{\alpha ^{D+1}},\quad
\alpha \ll 1  \label{sumn}
\end{equation}
we have
\begin{equation}
{\mathcal{I}}=\frac{\phi _{0}\Gamma (D+1)}{2^{D+1}\pi (1-x)^{D+1}}%
\int_{0}^{\infty }\frac{dz\,z^{D-2}}{\left( 1+z^{2}\right) ^{D/2+1}}=\frac{%
\phi _{0}(D-1)\Gamma ^{2}\left( \frac{D-1}{2}\right) }{16\pi (1-x)^{D+1}}%
,\quad \phi =0,\quad \phi =\phi _{0}.  \label{intsum5}
\end{equation}
Substituting this into formula (\ref{T00boundc}) to the leading order over $%
(a-r)^{-1}$one finds
\begin{equation}
\langle T_{00}\rangle _{c}\sim \frac{(4\xi -1)\Gamma \left( \frac{D+1%
}{2}\right) }{2^{D+2}\pi ^{(D+1)/2}(a-r)^{D+1}},\quad \phi =0,\phi
_{0},\quad r\rightarrow a.  \label{T00asra1}
\end{equation}
For $D=3$ and a conformally coupled scalar field this term coincides with
the first summand on the right of formula (\ref{18}) with the distance $%
(a-r) $ from the edge $r=a$, $\phi =0$ (or $\phi =\phi _{0}$) and
the opening angle $\pi /2$.

For the angles $0<\phi <\phi _{0}$ by using the formula
\begin{equation}
\sum_{n=1}^{\infty }n^{D}e^{-\alpha n}\cos n\beta =\frac{(-1)^{D}}{2}\frac{%
d^{D}}{d\alpha ^{D}}\left( \frac{\sinh \alpha }{\cosh \alpha -\cos \beta }%
-1\right) ,  \label{formser1}
\end{equation}
introducing a new integration variable $y=2\pi (1-x)\sqrt{1+z^{2}}/\phi _{0}$
and expanding over $(1-x)$ one finds that the leading contribution of the
term with $b^{(+)}(zx)\cos 2l\phi $ into (\ref{intsum1}) is equal to
\begin{equation}
\frac{\Gamma (D)}{2^{D+1}(1-x)^{D}}\left[ 1+\frac{1}{(D-1)(4\xi -1)}\right] .
\label{phitermas}
\end{equation}
In this case the leading contribution from the term with $b^{(-)}(zx)$
dominates and to the leading order by the way similar to that for (\ref
{T00asra1}) one has
\begin{equation}
\langle T_{00}\rangle _{c}\sim \frac{D(\xi -\xi _{c})\Gamma \left(
\frac{D+1}{2}\right) }{2^{D}\pi ^{(D+1)/2}(a-r)^{D+1}},\quad
0<\phi <\phi _{0},\quad r\rightarrow a.  \label{T00asra2}
\end{equation}
This leading divergence coincides with the corresponding one for a
cylindrical surface of the radius $a$ (see, for instance, \cite{saharianI}).

The surface divergences in the renormalised vacuum expectation values
of the local physical observables result from the idealization of
the boundaries as perfectly smooth surfaces which are perfect
reflectors at all frequencies, and are well known in quantum field
theory with boundaries. They are investigated in detail for
various types of fields and general shape of smooth boundary
\cite{Deutsch,kennedy}. Near the smooth boundary the leading
divergence varies as $(D+1)$th power of the distance from the
boundary. For conformally invariant fields the coefficient of this
leading term is zero. For non-smooth boundaries such as here, the
latter is not the case and this leads to the extra divergences in
the global quantities such as total Casimir energy (see, for
instance, \cite{nesterenko} for the case of a semi-circular
infinite cylinder and discussion in \cite{Dowk00}). It seems
plausible that such effects as surface roughness, or the
microstructure of the boundary on small scales (the atomic nature
of matter for the case of the electromagnetic field \cite{Cand80})
can introduce a physical cutoff needed to produce finite values of
surface quantities. An alternative mechanism for introducing a
cutoff which removes singular behavior on boundaries is to allow
the position of the boundary to undergo quantum fluctuations
\cite{Ford98}. Such fluctuations smear out the contribution of the
high frequency modes without the need to introduce an explicit
high frequency cutoff.

\begin{figure}[tbph]
\begin{center}
\begin{tabular}{ccc}
\epsfig{figure=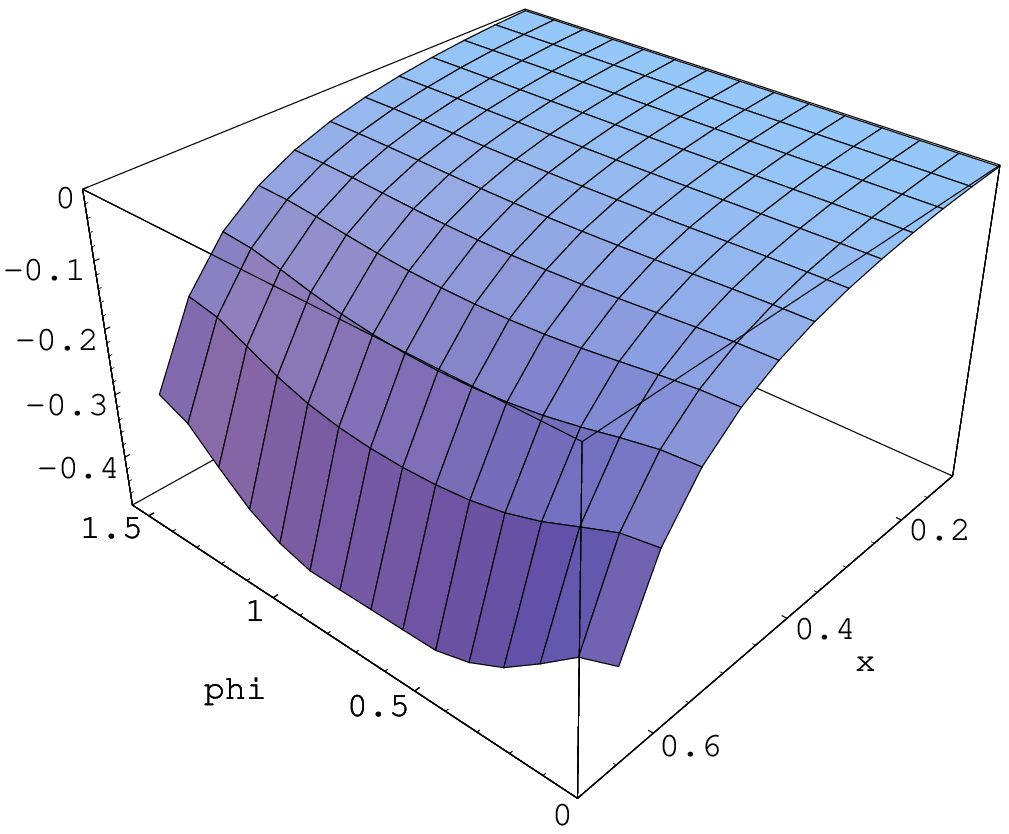,width=6cm,height=6cm} & \hspace*{0.5cm} & %
\epsfig{figure=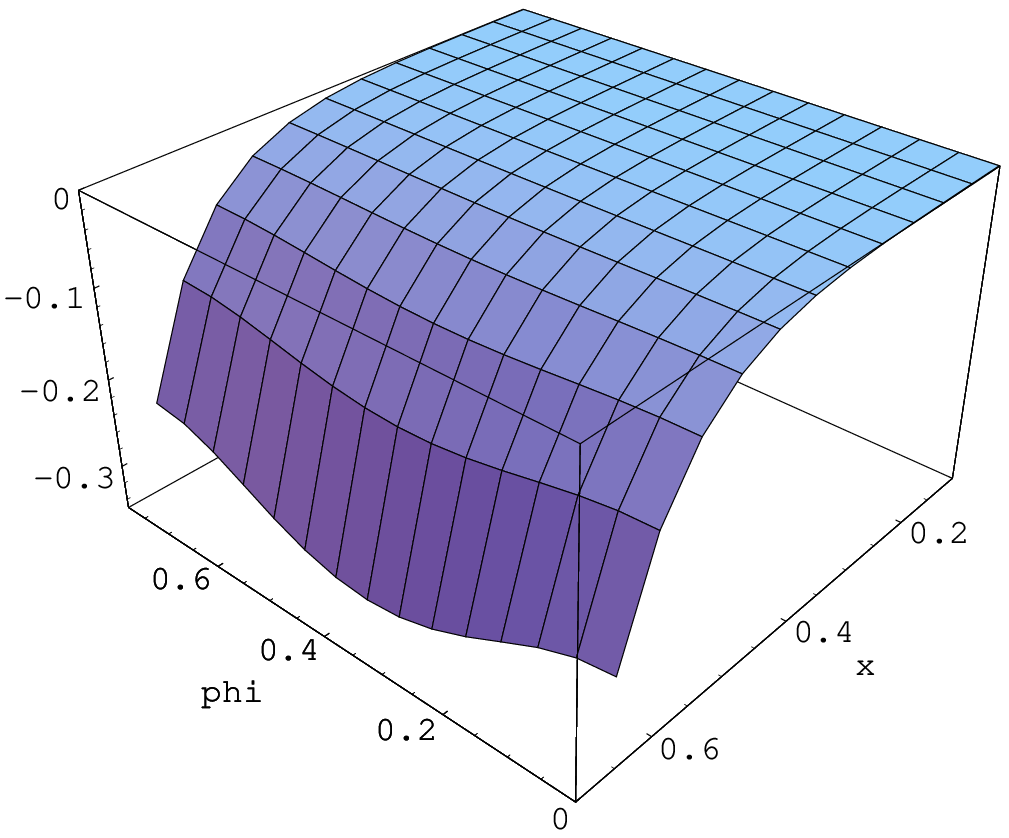,width=6cm,height=6cm}
\end{tabular}
\end{center}
\caption{ The expectation values for the cylindrical part of the
vacuum energy density, $a ^{D+1}\langle T_{00}\rangle _c$ for
$D=3$ minimally coupled scalar field in the cases $\phi _0=\pi /2$
(left) and $\phi _0=\pi /4$ (right), versus $\phi $ and $x=r/a$. }
\label{fig1min}
\end{figure}

\begin{figure}[tbph]
\begin{center}
\begin{tabular}{ccc}
\epsfig{figure=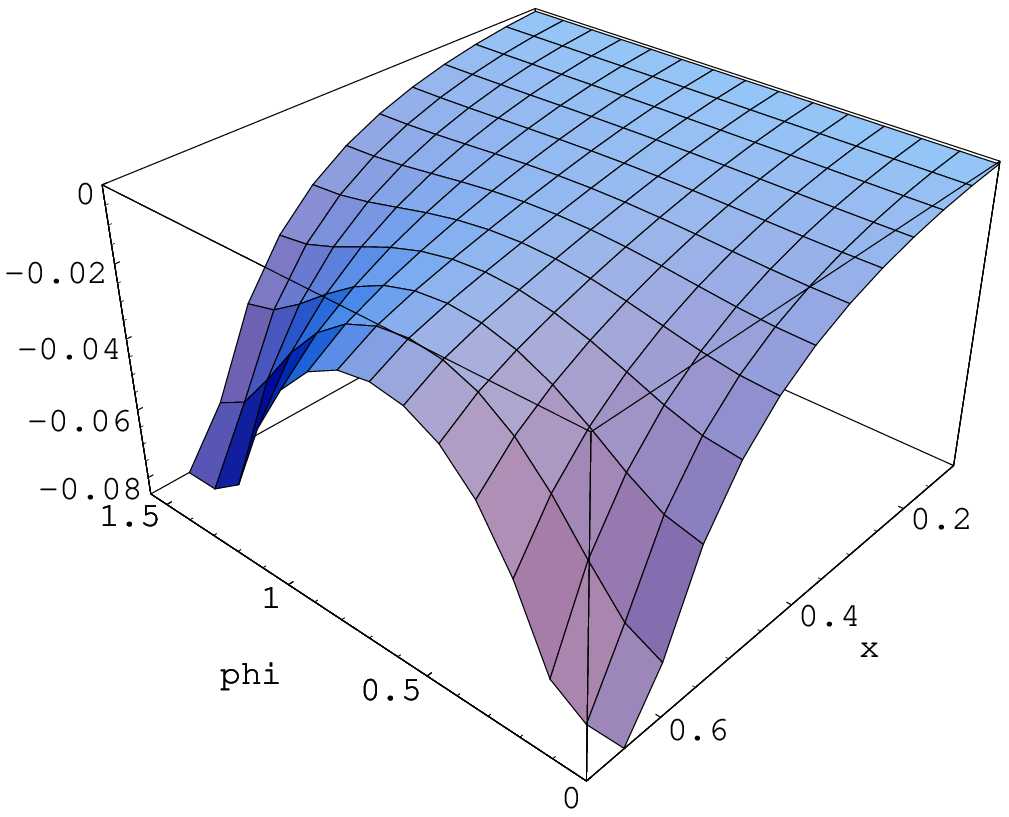,width=6cm,height=6cm} & \hspace*{0.5cm} & %
\epsfig{figure=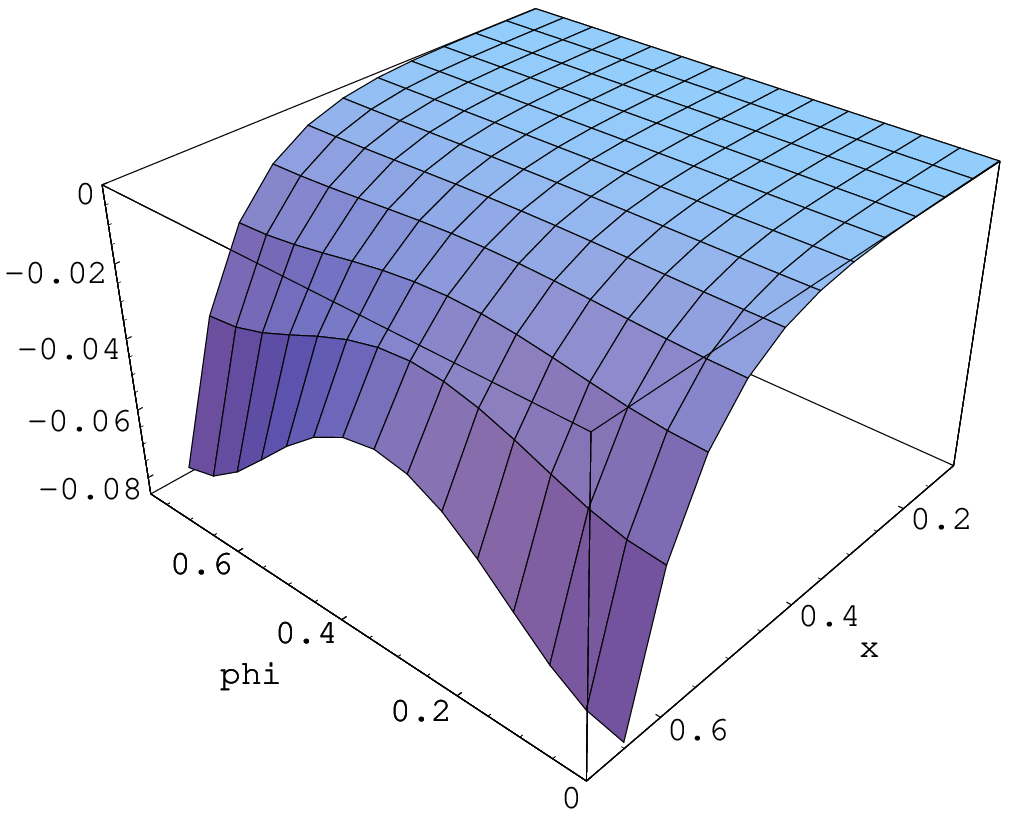,width=6cm,height=6cm}
\end{tabular}
\end{center}
\caption{ The same as in fig. \ref{fig1min} for a conformally
coupled scalar.} \label{fig2con}
\end{figure}

The dependence of the cylindrical boundary part of the vacuum
energy density multiplied by $a^{D+1}$, $a^{D+1}\langle
T_{00}\rangle _c$, on the angle $\phi $ and radial coordinate
$x=r/a$ is shown in figure \ref{fig1min} for a minimally coupled
scalar ($\xi =0$) and in figure \ref{fig2con} for a conformally
coupled scalar ($\xi =\xi _c$) in $D=3$. The cases $\phi _0=\pi
/2$ (left graphics) and $\phi _0=\pi /4$ (right graphics) are
presented, when the cylindrical part vanishes at the edge $r=0$.

\section{Final remarks} \label{sec:Conc}

In the present paper we calculate the vacuum energy density for a
massless scalar field with a general curvature coupling and
satisfying the Dirichlet boundary conditions  in the wedge--shaped
region formed by two plane boundaries intersecting at an arbitrary
angle $\phi _0$ with and without a cylindrical outer boundary. All
calculations are made at zero temperature and it is assumed that
the boundary conditions are frequency independent. The latter
means no dispersive effect is taken into account. The energy
density for the wedge, given by formula (\ref{18}), has been
obtained by standard application of the mode summation method.
This formula generalizes the result previously known in literature
for a conformally invariant scalar field. In the case of a
non-conformally coupled scalar the vacuum energy density is
angle-dependent and diverges on the wedge sides. For a minimally
coupled scalar it is negative everywhere inside a wedge. In the
case of presence of an outer cylindrical boundary, discussed in
section \ref{sec:Wedgecyl} the expectation value of the
corresponding vacuum energy density is presented in the form of
series over zeros of the Bessel function. The summation formula
for this type of series based on the generalized Abel-Plana
formula allows to extract explicitly from the expectation value
the part due to the wedge without a cylindrical shell. The
additional cylindrical contribution to the vacuum energy density
is presented in the form of the strongly convergent integrosum.
The latter is finite on the wedge sides but contains surface
divergences on the cylindrical boundary. At the edge $r=0$ it
vanishes for $0<\phi _0<\pi $, is finite for $\phi _0=\pi $, and
diverges as $r^{-2(1-\pi /\phi _0)}$ for $\phi _0>\pi $. The
generalization of the results obtained here for the Neumann, or
more general Robin boundary conditions is straightforward. For
instance, for the Neumann case in the expressions (\ref{eigfunc0})
and (\ref{eigfunccirc}) of the eigenfunctions the function $\cos
(l\phi )$ stands instead of $\sin (l\phi )$ and the quantum number
$n$ takes the values $0,1,2,\ldots $. In the case with a circular
outer boundary now the eigenvalues for $\gamma a$ are zeros for
the derivative of the Bessel function. The formula to sum the
series over these zeros can be taken from \cite{Saha87}.

\section*{Acknowledgments}

One of the authors (AHR) would like to thank Dr. Niels Walet and
Prof. R. F. Bishop for their supportive encouragement and
acknowledges support from ORS award. The work of AAS was supported
in part by the Armenian Ministry of Education and Science (Grant
No. 0887).

\end{document}